# Cascade hot carriers via broad-band resonant tunneling

Kamal Kumar Paul[1], Ashok Mondal[1,2], Jae Woo Kim[1,2], Ji-Hee Kim[1,2,*], and Young Hee Lee[1,2,3,*]

[1]Center for Integrated Nanostructure Physics, Institute for Basic Science, Sungkyunkwan University, Suwon, Republic of Korea

[2]Department of Energy Science, Sungkyunkwan University, Suwon 16419, Republic of Korea

[3]Department of Physics, Sungkyunkwan University, Suwon 16419, Republic of Korea

*kimj@skku.edu, leeyoung@skku.edu

**Extraction of hot carriers (HCs) over the band-edge is a key to harvest solar energy beyond Shockley-Queisser limit[1]. Graphene is known as a HC-layered material due to phonon bottleneck effect near Dirac point, but limited by low photocarrier density[2]. Graphene/transition metal dichalcogenide (TMD) heterostructures circumvent this issue by ultrafast carrier transfer from TMD to graphene[2,3]. Nevertheless, efficient extraction of photocurrent by means of HCs together with carrier multiplication (CM) is still missing. Here, we introduce an ultrathin broadband resonant tunneling (BRT) barrier, $TiO_X$ to efficiently extract photocurrent with simultaneous CM and HC measurements in $MoS_2$/graphene/$TiO_X$ heterostructure. The BRT layer gives rise to boosting open circuit voltage which is linearly proportional to incident photon energy. Meanwhile, short circuit current rises rapidly over $2E_g$ with obvious CM feature. This was explained by defining the joint density of states between graphene and $TiO_X$ layer over positive and negative voltage. The broadband resonant tunneling states inherently constructed from oxidation states varying from $Ti^{3+}$ to $Ti^{4+}$ allow the ultrafast HCs to efficiently transfer from graphene to $TiO_X$ layer. We find that the number of available tunneling states is directly proportional to short circuit current, which is well corroborated with $TiO_X$ and $MoS_2$ thickness variance. We obtained an optimum thickness of BRT layer of ~2.8 nm, yielding cascade open circuit voltage as high as ~0.7 V, two orders of magnitude higher than that without BRT layer to reach a record efficiency of 5.3% with improved fill factor owing to synergistic HC and CM conversion under 1-SUN with long-term stability.**

Manipulation of relaxation and extraction of the photo-excited carriers is one of the key factors for efficient hot carrier-driven photocatalytic, optoelectronic and photovoltaic applications[1,4,5]. The photo-generated carriers beyond the band-edge, often referred to as hot carriers (HCs), lead to heat dissipation through carrier-phonon scattering and finally arrive at the band-edge to contribute to photocurrent. This major intrinsic energy loss limits a maximum theoretical power conversion efficiency (PCE) of ~34% in a conventional single-junction solar cell under 1-SUN illumination[6]. HC-based solar cell essentially utilizes the excess energy of HCs to generate additional free carriers or free energy. This leads to enhanced short circuit current ($J_{SC}$) in carrier multiplication (CM) solar cell with a maximum PCE of ~46%[7,8] and improved open circuit voltage ($V_{OC}$) in HC solar cell with a maximum PCE of ~67% beyond Shockley-Queisser limit[9,10]. Efficient extraction of HCs to proliferate PCE is prerequisite to retain prolonged relaxation and more importantly, to extract HCs to the contact, often known as energy selective contact[11].

Numerous bulk materials including GaAs[12] and InN[13] have been proposed as a HC absorber with high carrier mobility and phononic bandgap. Extraction of HCs has been inefficient so far, however, primarily due to poorly implemented phonon bottleneck effect and energy selective contact[1,14]. Low-dimensional confined systems such as quantum dots (QDs) and van der Waals (vdWs) layered transition metal dichalcogenides (TMDs) predominantly involve strong Coulomb interaction and thus improve CM[8,15–20]. Nevertheless, QD structures via CM have been reported with inefficient carrier extraction due to the charge localization, low carrier mobility and diffusion length[21–24]. In particular, one unique feature in graphene (Gr) is linear band dispersion, where the carrier density vanishes near the Dirac point to persist in prolonged lifetime, resulting in strong phonon bottleneck effect at low energy of <200 meV near the Dirac point, i.e., hence HC effect[5,25–28]. While evidence of HCs has been demonstrated in Gr with clear phonon bottleneck effect, $V_{OC}$ is still limited to a few mV and furthermore, Gr possesses considerably low optical absorbance (~2.3% / monolayer) compared to that of vdW TMD materials ($MoS_2$, $MoSe_2$, $WS_2$, etc.) which exhibits relatively high absorbance (~5-10% / monolayer)[2,5]. An atomically thin $MoS_2$/Gr heterostructure demonstrates efficient nondissipative interfacial HC transfer from $MoS_2$ to Gr using ultrafast transient absorption spectroscopy, thereby increasing the photogenerated carrier population. Despite the promising conjectures[2,3,29], the design of Gr-based TMD architecture to improve amenable solar cell efficiency involving HCs and CM is still challenging[1].

In this article, we propose an ultrathin $TiO_X$ layer for broadband resonant tunneling to cascade open circuit voltage for HCs in $MoS_2$/Gr heterostructure device. The Gr layer is sandwiched between multilayer $MoS_2$ and $TiO_X$ layer. We introduce the joint density of states between Gr and $TiO_X$ layer to explain the broadband tunneling states in $TiO_X$ layer which is directly correlated to $V_{OC}$ in HC to further boost $J_{SC}$ in CM. Our report demonstrates the first proof-of-concept of HC solar cell based on vdWs $MoS_2$/Gr heterostructure boosted by the ultrafast transfer of multiplicated HCs from $MoS_2$ to Gr[2] and predominant phonon bottleneck effect in Gr, followed by the broadband resonant tunneling (BRT) through ultrathin BRT layer ($TiO_X$).

## Direct photocurrent measurement for carrier multiplication

Figure 1a shows the schematic of CM in $MoS_2$ with the optical excitation above $2E_g$ (process 1) and inverse Auger recombination (process 2). The vdW TMD materials are known to manifest low CM threshold energy of nearly two times the bandgap and ideal (~99%) CM conversion efficiency due to strong Coulomb interaction and quantum confinement effect using transient absorption spectroscopy[8]. Photoexcited carriers generate abundant HCs in $MoS_2$ and are ultrafast-transferred to Gr to become hotter due to phonon bottleneck effect and finally extract carriers to Ti electrode (Fig. 1b)[2]. Despite the presence of rich HCs in Gr, they become cold due to similar work function between Gr and Ti, leaving negligible $V_{OC}$[30]. Therefore, robust design of Gr-based TMD architecture is prerequisite. The reported vertical/ lateral structures have been mostly utilized by the Schottky barrier/ asymmetric electrodes and/or doping (chemically/ electrostatically) without considering the HCs (see Supplementary Table 1). We propose ultrathin $TiO_X$ layer as a broadband resonant tunneling barrier (Fig. 1c). $TiO_X$ layer is deposited onto Gr to efficiently extract multiplicated HCs to elevate $V_{OC}$ as well as $J_{SC}$. The $TiO_X$ layer plays a role as tunneling from cascade HC energies resonant by the presence of wide range of oxidation states, i.e., broadband resonant tunneling. Formation of the ultrathin $TiO_X$ layer is the key to create the broadband tunneling states ranging from $Ti^{3+}$ to $Ti^{4+}$ oxidation states so that HCs can be extracted with high $V_{OC}$. Furthermore, optimum thickness for tunneling is required to maximize the current density. This is conceptually distinguished from energy selective contact in bulk.[11,31]

Figure 1d shows current-voltage characteristics of $MoS_2$/Gr/Ti cell with various laser energies at a fixed absorbed photon density (Supplementary Fig. 1). $V_{OC}$ is negligibly small with ~ 10 mV for excitation laser energies up to 3.31 eV, which is still limited for HC extraction. Meanwhile, the photocurrent density is linearly proportional to the absorbed photon density up to ~$1.4\times10^{15}$ $cm^{-2}$ $s^{-1}$ with various excited energies where the absorbed photon density is multiplied by absorbance of the device (1.53 eV – 3.31 eV (Fig. 1e and see Extended Data Fig. 1 with the corresponding absorbance values in Supplementary Table 2). This excludes possibility of many-body interaction[32]. The linear slope is still retained with abrupt slope change beyond 2.54 eV ($2E_g$). The quantum yield can be defined by the slope at a given excitation photon energy, QY = $S_E/S_{2Eg}$ similar to previous report[32]. Figure 1f shows a clear CM feature in $MoS_2$ from $MoS_2$/Gr cell from photocurrent measurement. The quantum yield abruptly rises above $2E_g$ and approaches to as high as 200% at $3E_g$. Meanwhile, it is remarkable to observe an escalated $V_{OC}$ of ~0.7 V by two orders of magnitude in $MoS_2$/Gr/$TiO_X$ cell (Fig. 1g). Furthermore, $J_{SC}$ is enhanced to 5.7 mA $cm^{-2}$ by 14 times with enhanced quantum efficiency (Fig. 1h, i), similar to that of $MoS_2$/Gr cell.

## Cascade hot carriers to elevate $V_{OC}$ in $MoS_2$/Gr/$TiO_X$ heterostructure

Despite the similar CM conversion efficiency regardless of $TiO_X$ inclusion in the cell, $V_{OC}$ is elevated by two orders of magnitude by depositing $TiO_X$ layer on Gr by atomic layer deposition, where the $MoS_2$/Gr heterojunction is constructed by vdWs assembly of individual material on hBN substrate (Fig. 2a,b) (See Methods and Supplementary Fig. 2). To understand the underlying mechanism, we investigated the photocells with various photon densities and energies as well as $TiO_X$ layer thicknesses. Figure 2c presents the representative I-V curves by illuminating various photon densities at different photon energies (3.31 eV, 1.94 eV and 1.53 eV) for a given $TiO_X$ thickness of 2.8 nm (See Supplementary Fig. 3 for other energies). Both $V_{OC}$ and $J_{SC}$ are remarkably enhanced at higher incident energy of 3.31 eV (> $2E_g$) compared to low energy of 1.94 eV or 1.53 eV (< $2E_g$). Figure 2d shows $V_{OC}$ with absorbed photon density for given photon energy.

$V_{OC}$ rises sharply at low fluence and saturates at higher fluence. Unlike conventional solar cell, owning the $V_{OC}$ insensitive to incident energy, a HC solar cell in principle possesses dynamic $V_{OC}$ varying with excitation energy (Fig. 2d). Unlike low $V_{OC}$ of ~ 10 mV in $MoS_2$/Gr/Ti cell, the cascade $V_{OC}$ as high as 0.7 V is clearly manifested in $MoS_2$/Gr/$TiO_X$/Ti cell. $V_{OC}$ is linearly proportional to incident energy below $2E_g$ and saturates at higher energy (Fig. 2e) with no distinct threshold energy of $2E_g$.

Such a systematic $V_{OC}$ enhancement ensures the efficient extraction of HCs in the presence of $TiO_X$ layer. Gr/$TiO_X$/Ti cell without $MoS_2$ still exhibits comparatively amenable photovoltaic characteristics due to HC extraction with medium $V_{OC}$, indicating importance of the $TiO_X$ layer (See Extended Data Fig. 2). This is confirmed again by negligibly low $V_{OC}$ in $MoS_2$/Gr/Ti cell without $TiO_X$ layer caused by the well-known photothermal effect, similar to previous report in the Gr-metal interfaces[5]. The $J_{SC}$ increases linearly with incident energy below $2E_g$ and rises rapidly at > $2E_g$ in $MoS_2$/Gr/$TiO_X$ cell (Fig. 2f), indicating the CM effect similar to Figure 1i. Similar CM effect was observed in $MoS_2$/Gr/Ti cell without $TiO_X$ layer in Fig. 1f. Such a CM feature is not clearly visible in the absence of $MoS_2$ in Gr/$TiO_X$ cell, still retaining amenable $J_{SC}$, where the reasonable $V_{OC}$ in $TiO_X$ layer still contributes to $J_{SC}$. $MoS_2$ generates additional carriers via CM but fails to extract carriers to Ti layer (low work function difference) in $MoS_2$/Gr/Ti cell. In Gr/$TiO_X$ cell, both CM and HCs are provoked in Gr and efficiently extracted through $TiO_X$ BRT barrier to reveal reasonable $J_{SC}$. In ideal $MoS_2$/Gr/$TiO_X$ cell, both CM from $MoS_2$ and Gr, and efficient HC extraction though $TiO_X$ layer enhances both $J_{SC}$ and $V_{OC}$.

## Optimum thicknesses of $TiO_X$ and $MoS_2$ layer

Figure 3 shows the photovoltaic characteristics on the BRT layer and $MoS_2$ film thicknesses for a given incident energy of 3.31 eV. At ~1 nm thin $TiO_X$ layer, the leakage current is dominant from negative voltage and hence $V_{OC}$ remains small (Fig. 3a). As the $TiO_X$ layer becomes thicker, the leakage current seemingly becomes smaller. Both $V_{OC}$ and $J_{SC}$ are enhanced particularly at 2.8 nm and are degraded at thicker layers (see Supplementary Fig. 4). $V_{OC}$ increases as a function of photon flux to saturate at higher photon density (Fig. 3b) owing to band filling effect[32,33], which reaches the maximum at a $TiO_X$ layer of 2.8 nm. Similarly, the $J_{SC}$ is maximum at a $TiO_X$ layer of 2.8 nm (Fig. 3c). The maximum $V_{OC}$ and $J_{SC}$ are clearly observed at an optimum $TiO_X$ layer thickness of 2.8 nm (Fig. 3d,e). At ultrathin $TiO_X$ layer i.e., 1 nm, the leakage is dominant and $V_{OC}$ is not matured yet, limiting $J_{SC}$. At an optimum $TiO_X$ layer of 2.8 nm, the leakage current is minimized, resulting in maximum $J_{SC}$. At a thick $TiO_X$ layer over 2.8 nm, the $J_{SC}$ is degraded. Similarly, $V_{OC}$ rises with increasing $TiO_X$ layer. At thick $TiO_X$ layer, the surface states are relatively reduced with remaining bulk states to consequently reduce $V_{OC}$.

Figure 3f presents the I-V characteristics with various $MoS_2$ film thicknesses (1.4 – 20 nm) at a fixed BRT layer thickness of 2.8 nm with 3.31 eV (details shown in Supplementary Fig. 5). We plot $V_{OC}$ (Fig. 3g) and $J_{SC}$ (Fig. 3h) with absorbed photon density, revealing the systematic uptakes with the $MoS_2$ film thicknesses. $J_{SC}$ is linearly proportional at thin $MoS_2$ layer and saturates at thick $MoS_2$ layer (Fig. 3i). This can be fitted to $J = J_0 - J^{max} e^{-x/\lambda}$, where $J_0 = J_{Gr} + J^{max}$, $J_{Gr}$ is the maximum current density value for Gr/$TiO_X$ cell without $MoS_2$, and $J^{max}$ is the current density in the presence of $MoS_2$. The characteristic diffusion length of carrier, λ is estimated as ~6.5 nm. Thick layer over λ, i.e., ~30 nm reaches maximum $J_{SC}$ due to diffusion limit. Thicker $MoS_2$ absorbs more excited carriers due to enhanced absorbance to further proliferate more excited carriers via CM. These are further transferred to Gr layer to become hotter to contribute higher $V_{OC}$ owing to band filling effect (Fig. 3j).

**Manifestation of resonant tunneling states in $TiO_X$ layer**

So far, we observed the cascade $V_{OC}$ and $J_{SC}$ in terms of CM and HCs in $MoS_2$/Gr/$TiO_X$ heterostructure by introducing the BRT barrier with an ultrathin $TiO_X$ layer. To understand microscopic origin of BRT barrier, we analyzed oxidation states of the $TiO_X$ layer using XPS and the joint density of states (*dI/dV*) of the heterostructure cell. Figure 4a represents the schematic of broadband resonant tunneling states inherently constructed from oxidation states in $TiO_X$ layer varying with $Ti^{3+\delta}$. Ultrathin $TiO_X$ layer with various oxidation states ($Ti^{3+\delta}$, δ = 0 to 1) forms a broadband tunneling states and thus enables a cascade tunneling resonant with energy of the multiplicated hot electrons in Gr, while completely blocking the holes (Fig. 4a). Existence of the different oxidation states in BRT barrier is confirmed by O1s (Fig. 4b) and Ti2p (Fig. 4c) core-level XPS spectra, identifying $Ti^{3+}$ and $Ti^{4+}$ oxidation states[34,35]. Both surface states and bulk states may vary under different environmental conditions including structural deformation in $TiO_X$ lattice, hence defined as $Ti^{3+\delta}$, δ = 0 to 1.

Figure 4d represents the joint density of states *dI/dV* between Gr and $TiO_X$ layer with various $TiO_X$ layer thicknesses. Each curve is color-filled up to 1$^{st}$ peak minimum, which is equivalent to $V_{OC}$. At thick $TiO_X$ layer of 6 nm, the bulk states prevail with relatively low $V_{OC}$. At lower $TiO_X$ layer of 5 nm, the surface states start to emerge and hence increase $V_{OC}$, while the bulk states shrink. The surface states become maximum at $TiO_X$ layer of 2.8 nm, eventually maximizing $V_{OC}$, while still retaining amenable bulk states. The number of available tunneling states can be obtained by integrating *dI/dV* over the positive voltage,

$$N_{TS} = \int_{V=0}^{Voc} (dI/dV)_V \, dV. \quad (1)$$

$N_{TS}$ is maximum at $TiO_X$ layer of 2.8 nm (Fig. 4e), similar to $V_{OC}$ and $J_{SC}$. Such a trend in $TiO_X$ layer thickness is different from that in $MoS_2$ channel thickness. $N_{TS}$ seemingly rises with increasing $MoS_2$ layer thickness (Fig. 4f) due to multiplicated carriers in $MoS_2$ layer to transfer to Gr, eventually contributing to $J_{SC}$ through BRT barrier. The $N_{TS}$ in Figure 4g is congruent with $J_{SC}$ behavior observed in Figure 3i.

The number of leakage states, $N_{LS}$, can be similarly defined by integration of *dI/dV* over the negative voltage,

$$N_{LS} = \int_{V=0}^{-V} (dI/dV)_V \, dV. \quad (2)$$

Figure 4h summarizes $N_{LS}$ in terms of $MoS_2$ thickness (See Extended Data Fig. 3a). The leakage current is provoked by the presence of traps at $MoS_2$/Gr interface during ultrafast photoexcited carrier transfer from $MoS_2$ to Gr layer. At 1.4 nm thin $MoS_2$ layer, $N_{LS}$ prevails by the ample surface states of $MoS_2$ layer and eventually saturates by minimizing the trap density at thicker $MoS_2$ layer. Such a reduction of $N_{LS}$ with increasing $MoS_2$ layer thickness is well understood by inversely proportional $J_{SC}$. Similarly, $N_{LS}$ prevails at 1 nm thin $TiO_X$ layer due to the dominant back scattering of premature $V_{OC}$ (Extended Data Fig. 3b,c). The $N_{LS}$ rapidly drops at 2.8 nm thin TiOx layer due to ample tunneling states of mature $V_{OC}$. This further saturates at thicker $TiO_X$ layers in accordance with the degraded $V_{OC}$ or $J_{SC}$. Analysis of $N_{LS}$ is complementary to understand the photovoltaic performance together with $V_{OC}$ and $J_{SC}$.

## Solar cell performance of various heterostructures under 1-SUN illumination

Figure 5a depicts photocurrent density vs. voltage of different cells $MoS_2$/Gr/Ti, Gr/$TiO_X$/Ti, $MoS_2$/Gr/$TiO_X$/Ti and $MoS_2$/Gr/$TiO_X$ under 1-SUN illumination (AM 1.5 G). We observe remarkable cell performance for $MoS_2$/Gr/$TiO_X$/Ti with improved $V_{OC}$ and $J_{SC}$ compared to $MoS_2$/Gr/Ti or Gr/$TiO_X$/Ti cell. Such superb photovoltaic performance in $MoS_2$/Gr/$TiO_X$/Ti (Fig. 5b) is well explained by the significantly reduced number of leakage states at negative voltage (Extended Data Fig. 4a). This again reassures that such an analysis of $N_{TS}$ (Extended Data Fig. 4b) at positive voltage and $N_{LS}$ at negative voltage is inherent to explain origin of $V_{OC}$ and $J_{SC}$ under 1-SUN illumination.

The photovoltaic performance is enhanced further in $MoS_2$/Gr/$TiO_X$ cell under equivalent 1-SUN illumination (considering the transmission loss due to the top Ti electrode). The fill factor (FF) reaches as large as 63.4% in $MoS_2$/Gr/$TiO_X$ cell (Fig. 5c), similar to other materials such as perovskites[36,37] or Si/perovskite tandem solar cells[38], over two times larger than in $MoS_2$/Gr cell due to inherent construction of the BRT barrier with optimized $TiO_X$ layer thickness. The HC-dominant Gr/$TiO_X$/Ti cell outperforms the CM-dominant $MoS_2$/Gr/Ti cell by two orders of magnitude higher PCE (Fig. 5c). Surprisingly, we observe the improved PCE of 5.3% in $MoS_2$/Gr/$TiO_X$ cell via synergistic effects of HCs and CM. Efficient HC extraction with synergistic CM in TMD materials by introducing ultrathin BRT layer has not been observed previously to date. The extracted electrical power density ($P_{el}$ = photocurrent density × voltage) versus voltage is shown for different cells (Fig. 5c, inset). The maximum electrical output power density ($P_{el,m}$) for $MoS_2$/Gr/$TiO_X$/Ti cell is 3.5 mW $cm^{-2}$ at 0.52 V, which is further enhanced to 5.3 mW $cm^{-2}$ at 0.54 V for $MoS_2$/Gr/$TiO_X$ cell with three orders of magnitude higher than $MoS_2$/Gr/Ti cell ($P_{el,m}$ = 0.001 mA $cm^{-2}$ at 0.006 V). Figure 5d further demonstrates the long-term operational stability of $MoS_2$/Gr/$TiO_X$ cell under regular laboratory environment (see Methods), by revealing the persistent PCE with negligible fall even after 120 days, which establishes the stability of our HC-solar cell in TMD materials.

## Conclusion

We have introduced an ultrathin broadband resonant tunneling (BRT) barrier, $TiO_X$ to boost open circuit voltage and further improve photocurrent extraction from simultaneous measurements of CM and HC in $MoS_2$/Gr/$TiO_X$ heterostructure. Ultrafast-transferred multicarriers from $MoS_2$ enrich HCs in Gr due to phonon bottleneck effect. The cascade HC energies in Gr resonant with the broadband oxidation states in ultrathin $TiO_X$ layer ($Ti^{3+\delta}$, $\delta = 0$ to 1) are harvested efficiently via broadband resonant tunneling, elevating both $V_{OC}$ and $J_{SC}$. We were able to demonstrate the presence of broadband resonant tunneling states between Gr and TiOx layer by integrating joint density of states of *dI/dV*. The number of available tunneling states is directly proportional to short circuit current, which is well corroborated with $TiO_X$ and $MoS_2$ thickness variance. Although $MoS_2$/Gr/$TiO_X$ cell provides the PCE of ~5.3% with synergistic contribution of HCs and CM, this is just an infant stage to develop. There is plenty of room to revolutionize PCE improvement in TMD materials. Both optical and electronic designs need to be improved. The optical absorption of our vdWs heterojunction is only ~50%, which may be improved by replacing the TMD material or different device architecture[39]. The 5 nm thin top Ti contact layer is inevitably oxidized from its surface, thereby increasing the contact resistance. Moreover, the absorbed optical power in TMD/Gr heterostructure is weakened due to undesirable absorption in top Ti electrode, which could be rectified with an inverted cell architecture. Photovoltaic performance could be tuned further with suitable substitution of the BRT layer, or by engineering the BRT layer i.e., broadening the oxidation states, by increasing the available tunneling states and $V_{OC}$. Alternatively, different BRT barrier materials could be explored for optimizing PCE. Our proposal to analyzing resonant tunneling states using *dI/dV* and its integration at positive voltage and leakage current at negative voltage will provide deep insight to develop next-generation solar cells for other materials including perovskites and heterostructures.

## Methods

### Device fabrication

$MoS_2$ with various thicknesses (~1.4 − 20 nm), Gr (~2 nm), and hBN (~15 − 30 nm) were mechanically exfoliated from commercially available high-quality bulk crystals (2D Semiconductors, USA). To fabricate hBN/$MoS_2$/Gr vdWs heterostructures, the selected flakes were picked up and dry released on the Si/$SiO_2$ substrate by using PDMS/poly-propylene carbonate (PPC) stamp and a customized micro-manipulator. This technique excludes any interfacial contamination[40]. To achieve the desired geometry, the vertical stacks were shaped by the standard electron-beam (e-beam) lithography with 950 PMMA A4 e-beam resist, and sequentially etched by $O_2$ (for Gr) as well as $SF_6$ (for hBN, $MoS_2$) plasma using reactive ion etching (RIE). The bottom contact on Gr were defined by e-beam lithography, following the metal electrode Cr/Au (5/40 nm) deposition using the thermal evaporator. The ultrathin BRT layer with various thicknesses ($TiO_X$: 1 – 6 nm) was then deposited on the heterojunction by using atomic layer deposition (ALD)[41]. Finally, one window was constructed precisely on the $MoS_2$/Gr heterojunction for the top electrode by using e-beam lithography, followed by the deposition of thin Ti (5 nm) using ultra-high vacuum sputter (~ $10^{-7}$ torr) system.

## Characterization

The photophysical properties of the individual material as well as the vertical heterostructure were investigated by atomic force microscopy (Seiko AFM 5000 II) and optical spectroscopy (Raman) using NT-MDT with 532 nm laser excitation. X-ray photoelectron spectroscopy (K-Alpha, THERMO FISHER) was employed to identify the various oxidation states of $TiO_X$ layer. For absorption (A) measurements, reflection (R) and transmission (T) of the vdWs vertical stacks on transparent quartz substrate (transmission ~95%) were measured. Transmission spectra were recorded on the heterojunction considering the bare quartz substrate as a reference. The reflection spectra on the sample were measured after the calibration with a silver mirror (PF10-03-P01, Thorlabs) for the absolute reflection spectrum. Finally, the following equation was used for the absorption calculation: $A(\lambda) = 1 - R(\lambda) - T(\lambda)$.

**I-V measurements with lasers.** Photocurrent measurements of various vertical cells (Gr/$TiO_X$/Ti, $MoS_2$/Gr/Ti and $MoS_2$/Gr/$TiO_X$/Ti) were performed at the vacuum probe station (~$10^{-6}$ torr) using Keithley 4200A. External laser linings with various energies (diode lasers; 1.53 – 3.31 eV) were integrated with the vacuum probe station, and precisely incident on the sample through the optical window.

**1-SUN I-V measurements.** To determine the PCE of the vertical cells under AM 1.5 G, I-V measurements were performed using a solar simulator (Oriel Sol3A, Newport) connected with a Keithley 2400 source meter. The lamp intensity was calibrated with a Si reference cell (Newport, Oriel 91150 V) precisely placed at the sample position. I–V curves were recorded in a normal lab environment (temperature: 22 ºC; humidity: 35%). During the measurement, temperature of the sample was maintained via convection cooling with an electric fan. Photocurrent density as well as the PCE extracted from the continuous maximum power point tracking demonstrates the long-term operational stability of the cell under ambient conditions. I-V measurements were performed continuously for 3 hrs in each day and repeated for several times up to 120 days. In all I-V measurements, the voltage was applied to the Gr bottom contact, while the Ti top contact was grounded.

## Data availability

The data that support the findings of this study are available from the corresponding author upon reasonable request.

## Acknowledgements

This work was supported by the Institute for Basic Science of Korea (IBS-R011-D1) and Advanced Facility Center for Quantum Technology.

**Author contributions:** K.K.P. conceived the idea, and developed the project with the guidance of J.-H.K. and Y.H.L. K.K.P, J.-H.K. and Y.H.L. designed the experiments. K.K.P. fabricated all the devices with the help of A.M. K.K.P. performed all the optoelectronic and photovoltaic characterizations together with absorption, AFM and Raman measurements. J.W.K. contributed to the X-ray photoemission spectroscopy. K.K.P. and Y.H.L. interpreted the results, and wrote the manuscript. All authors discussed the results and commented on the manuscript. Y.H.L. supervised the entire project.

## Competing interests

The authors declare no competing interests.

## Additional information

**Supplementary information** The online version contains supplementary material available at

## Figures and captions

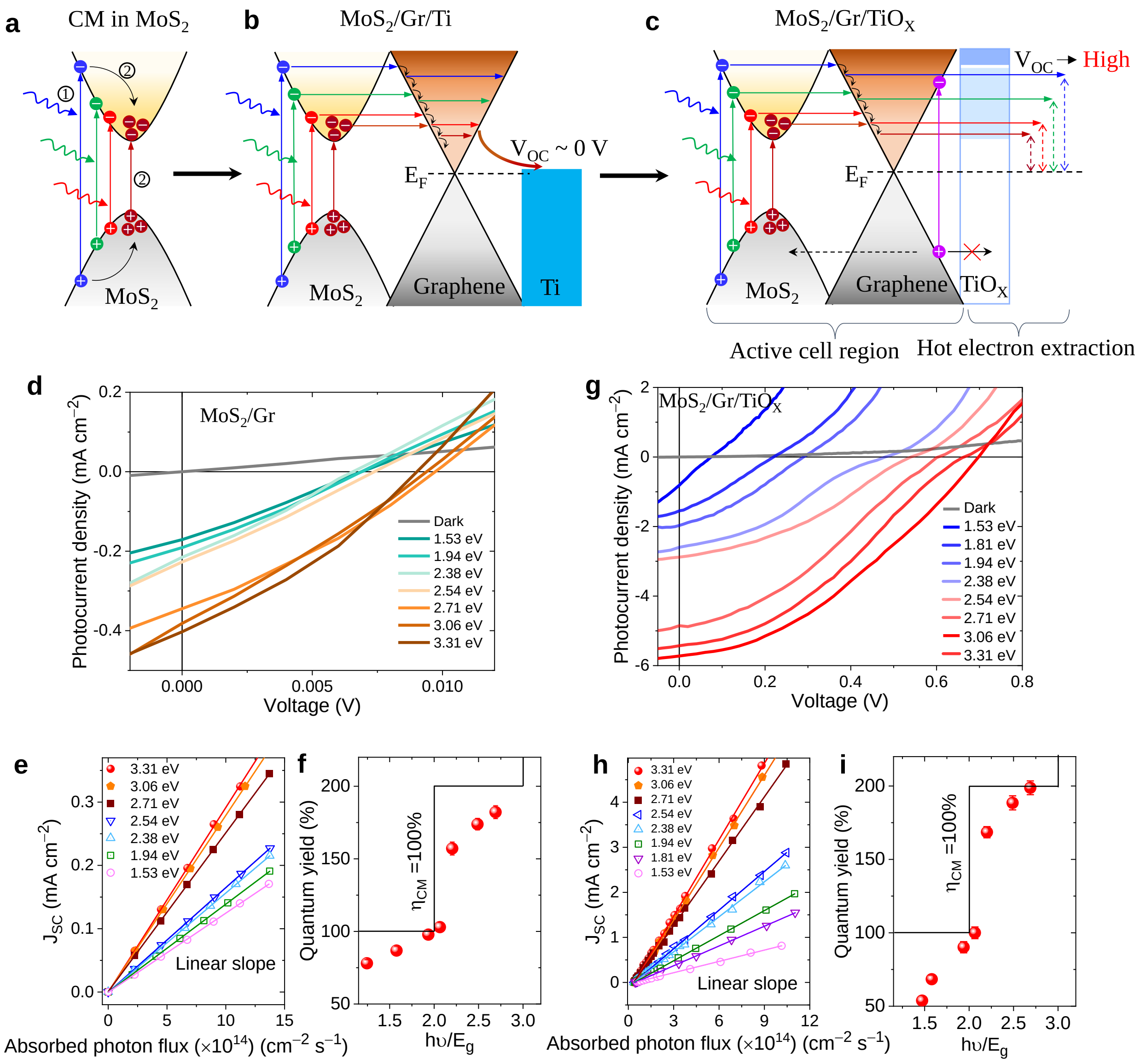

**Fig. 1| Carrier multiplication and hot carrier-driven photovoltaics in $MoS_2$/Gr/$TiO_X$ cell. a**, CM in $MoS_2$ with an optical excitation above $2E_g$ (process 1) and generated by inverse Auger recombination (process 2). **b**, Photoexcited multicarriers in $MoS_2$ is transferred to Gr which generates abundant HCs due to phonon bottleneck effect, finally reaching Ti electrode. Negligible $V_{OC}$ is detected between Gr and Ti electrode. **c**, $TiO_X$ is deposited on Gr for broadband resonant tunneling layer to efficiently extract HCs to elevate $V_{OC}$. **d**, Current-voltage characteristics of $MoS_2$/Gr cell with the excitation of various energies at a fixed absorbed photon density (~$1.4\times10^{15}$ $cm^{-2}$ $s^{-1}$). Maximum $V_{OC}$ and $J_{SC}$ are ~10 mV and 0.4 mA $cm^{-2}$ at 3.31 eV, respectively. **e**, Linear plot of $J_{SC}$ versus absorbed photon density upon optical excitations with various energies and **f**, CM conversion efficiency determined by the slope with respect to the bandgap of multilayer $MoS_2$

(1.23 eV)[42]. **g-i**, Similar characteristics of $MoS_2$/Gr/$TiO_X$ cell. Similar CM conversion efficiency is achieved but with much higher $V_{OC}$ (~0.7 eV) and improved $J_{SC}$ (~5.7 mA $cm^{-2}$).

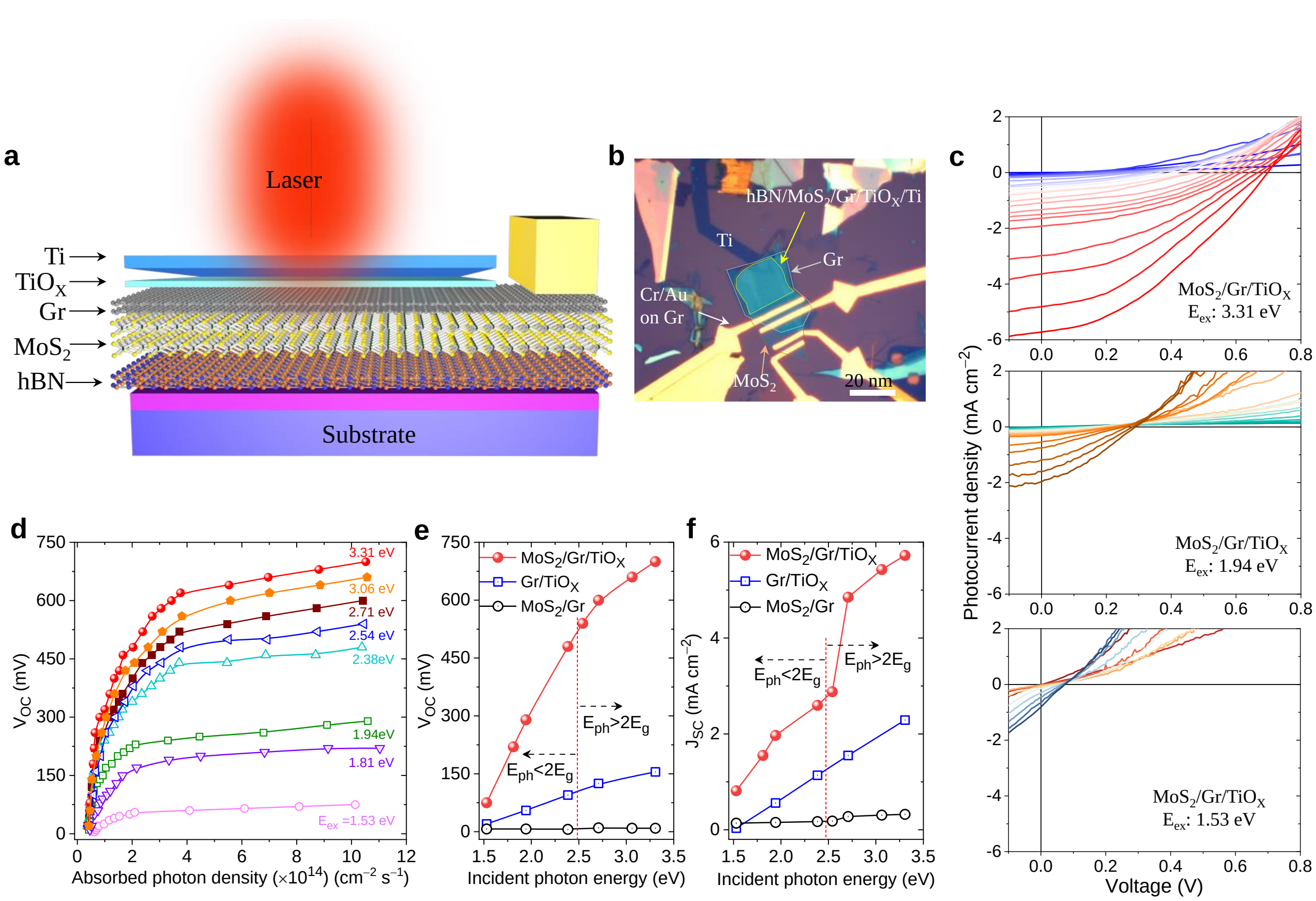

**Fig. 2| Hot carrier-driven photovoltaic $MoS_2$/Gr/$TiO_X$ cell.** **a**, Schematic illustration of $MoS_2$/Gr/$TiO_X$ cell with top Ti electrode (5 nm) fabricated on hBN substrate. **b**, Optical micrograph of the cell. **c**, Photocurrent density vs voltage with various photon densities (up to ~$1.1\times10^{15}$ $cm^{-2}$ $s^{-1}$) at different photon energies (3.31 eV, 1.94 eV and 1.53 eV). **d**, Open circuit voltage ($V_{OC}$) as a function of absorbed photon density for eight different energies. Variation of **e**, $V_{OC}$ as well as **f**, $J_{SC}$ with the incident photon energy at a fixed photon density ($1.1\times10^{15}$ $cm^{-2}$ $s^{-1}$) for different cells. $MoS_2$/Gr/$TiO_X$ cell shows cascade photovoltaic characteristics (high $V_{OC}$ and $J_{SC}$) compared to other cell architectures, confirming the coexistence of HC and CM.

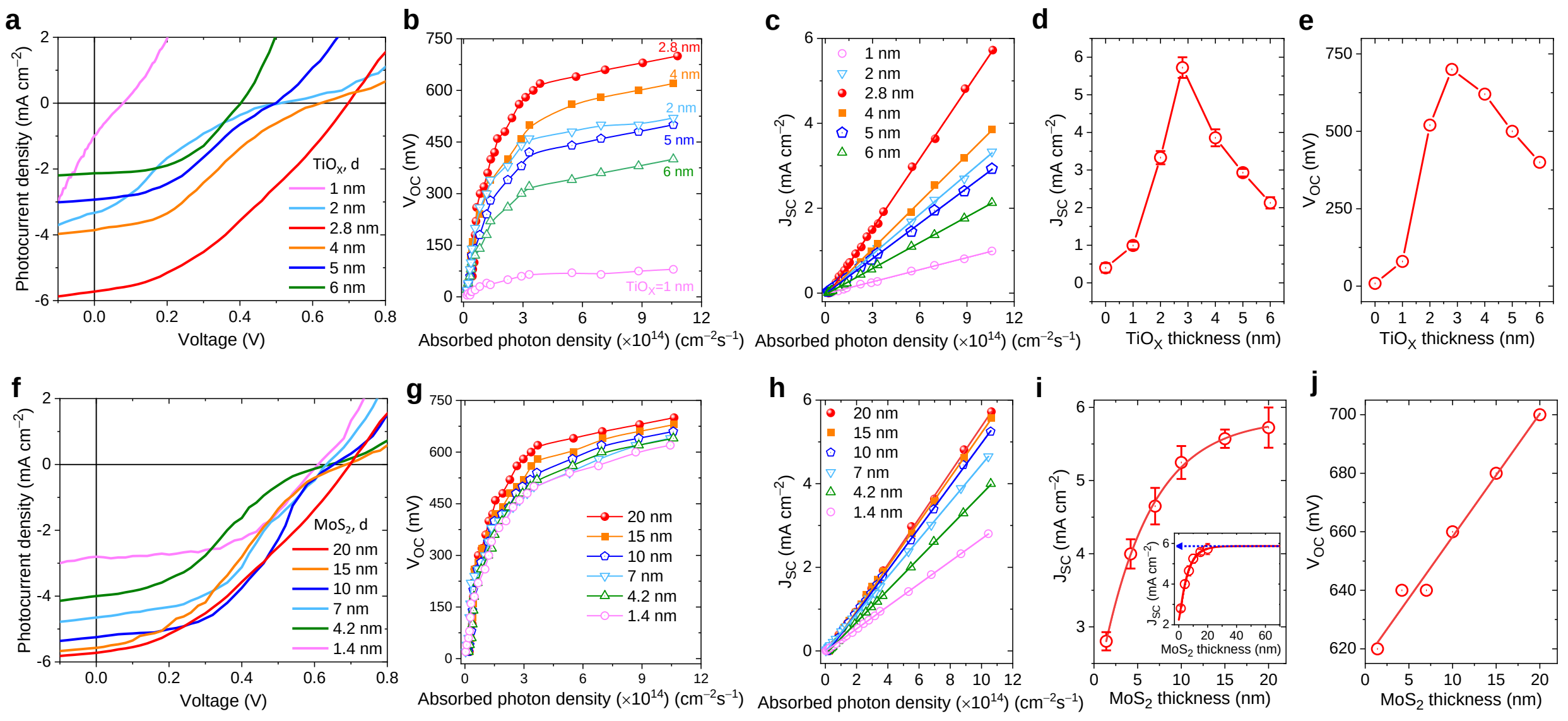

**Fig. 3| Thickness dependence of BRT barrier and $MoS_2$ film on photovoltaics. a**, Current-voltage characteristics of $MoS_2$/Gr/$TiO_X$/Ti cell with various BRT layer thicknesses ($TiO_X$: 1 ~ 6 nm) at 3.31 eV excitation. **b**, $V_{OC}$ and **c**, $J_{SC}$ as a function of absorbed photon density at 3.31 eV for different BRT layer thicknesses. Variation of **d**, $J_{SC}$ and **e**, $V_{OC}$ as a function of BRT layer thickness at 3.31 eV laser excitation and a fixed photon density (~$1.1\times10^{15}$ cm$^{-2}$ s$^{-1}$). The optimum $TiO_X$ layer thickness is determined as 2.8 nm with maximum $V_{OC}$ = 0.7 V and $J_{SC}$ = 5.7 mA cm$^{-2}$. **f**, Current-voltage characteristics of $MoS_2$/Gr/$TiO_X$/Ti cell with various $MoS_2$ film thicknesses (1.4 ~ 20 nm) at a fixed $TiO_X$ layer thickness of 2.8 nm under 3.31 eV excitation. **g**, $V_{OC}$ and **h**, $J_{SC}$ as a function of absorbed photon density at 3.31 eV for different $MoS_2$ film thicknesses. **i**, $J_{SC}$ increases linearly in thin $MoS_2$ layers and saturates at thick layers, while **j**, $V_{OC}$ increases linearly. Extrapolation of the fitting shows that the maximum $J_{SC}$ can reach as high as ~5.87 mA cm$^{-2}$ (inset of **i**).

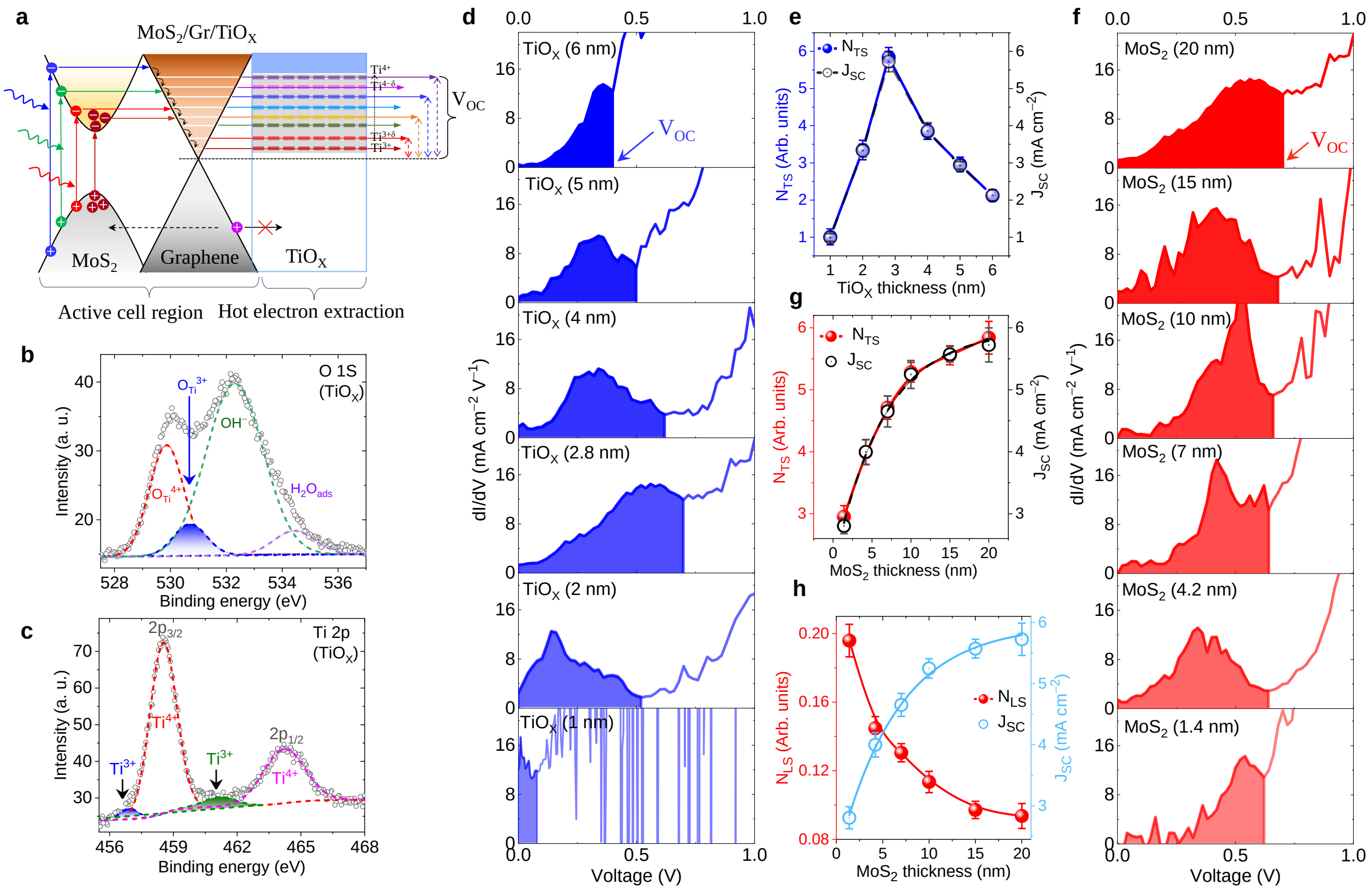

**Fig. 4| Origin of broadband resonant tunneling. a**, Cascade HC energies in Gr resonant with the various oxidation states in ultrathin BRT layer ($Ti^{3+\delta}$, $\delta$ = 0 to 1) is the key for efficient HC extraction through broadband tunneling. **b,** O1s and **c**, Ti2p core level XPS spectra of $TiO_X$ layer, fitted with Shirley baseline. Symbols represent the experimental data and the dashed lines correspond to the Gaussian fit. Fitted curves with filled area are associated with the various oxidation states of $Ti^{3+\delta}$, identifying the broadband resonant tunneling of HCs. Here, $\delta$ varies from 0 to 1. 1st derivative of photocurrent with respect to the voltage (*dI/dV*) with various **d**, BRT layer thickness and **f**, $MoS_2$ film thickness. Each peak shows the corresponding resonant tunneling through the BRT layer. Number of available tunneling states, $N_{TS}$ in the BRT layer with the variation of **e**, $TiO_X$ and **g**, $MoS_2$ thickness. **h**, Number of leakage states, $N_{LS}$ varying with $MoS_2$ thickness. Both the $N_{TS}$ and $N_{LS}$ are consistent with the corresponding $J_{SC}$ values.

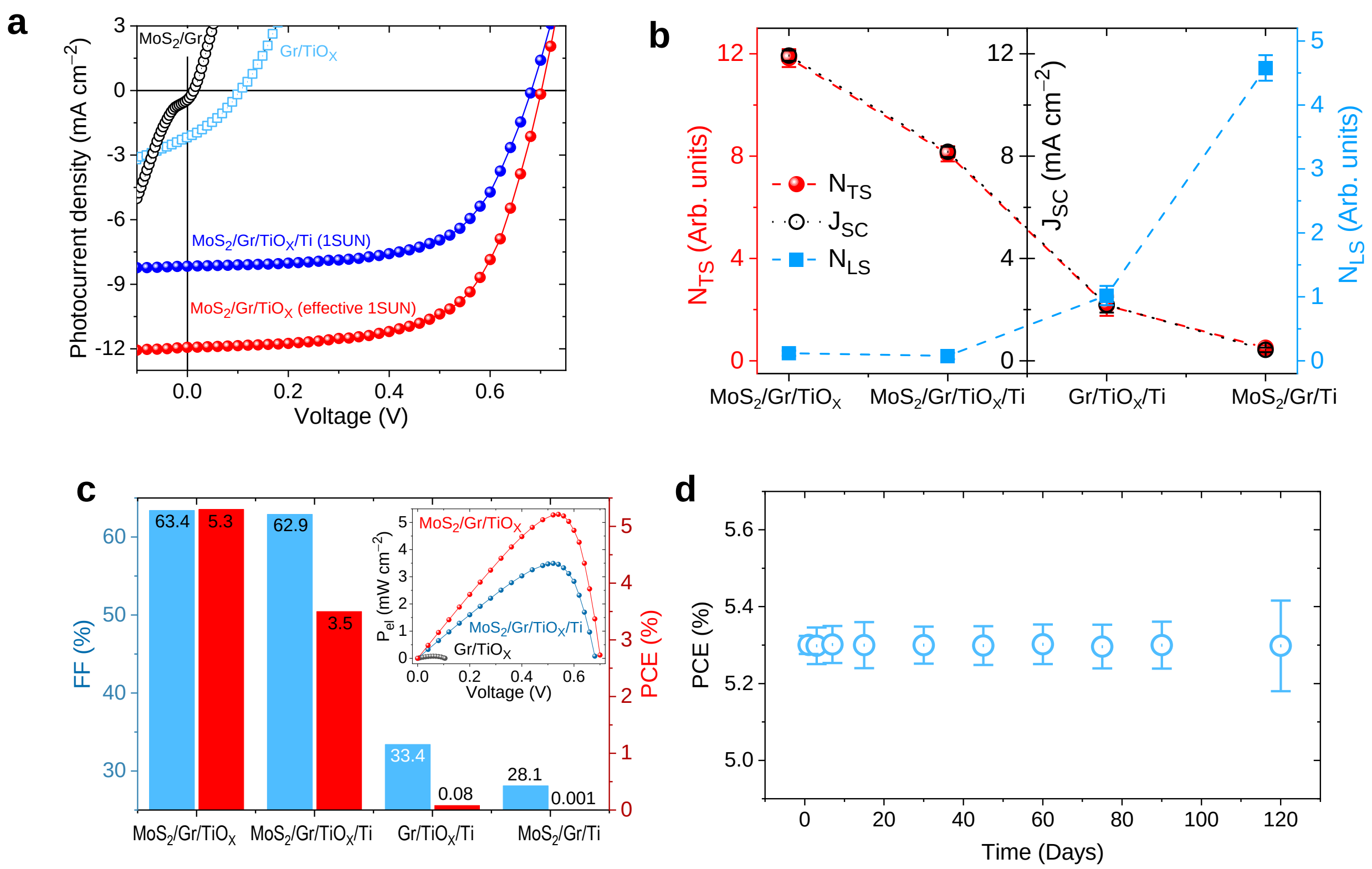

**Fig. 5| Solar cell performance in 1-SUN of $MoS_2$/Gr/$TiO_X$ cell. a**, Measured photocurrent density vs. voltage for different cells ($MoS_2$/Gr/Ti, Gr/$TiO_X$/Ti, $MoS_2$/Gr/$TiO_X$/Ti and $MoS_2$/Gr/$TiO_X$). Transmittance of Ti layer (5 nm) is ~67%. 1-SUN of $MoS_2$/Gr/$TiO_X$ is equivalent to 1.5-SUN of $MoS_2$/Gr/$TiO_X$/Ti. **b**, Number of available tunneling states ($N_{TS}$) at positive voltage and leakage states ($N_{LS}$) at negative voltage with corresponding $J_{SC}$ values. **c**, Comparison of fill factor (FF), electric power density ($P_{el}$, inset), and power conversion efficiency (PCE) of various cells. $MoS_2$/Gr/$TiO_X$ cell shows the maximum photovoltaic performance with PCE of 5.3%. **d**, Long-term operational stability of $MoS_2$/Gr/$TiO_X$ cell under ambient conditions.

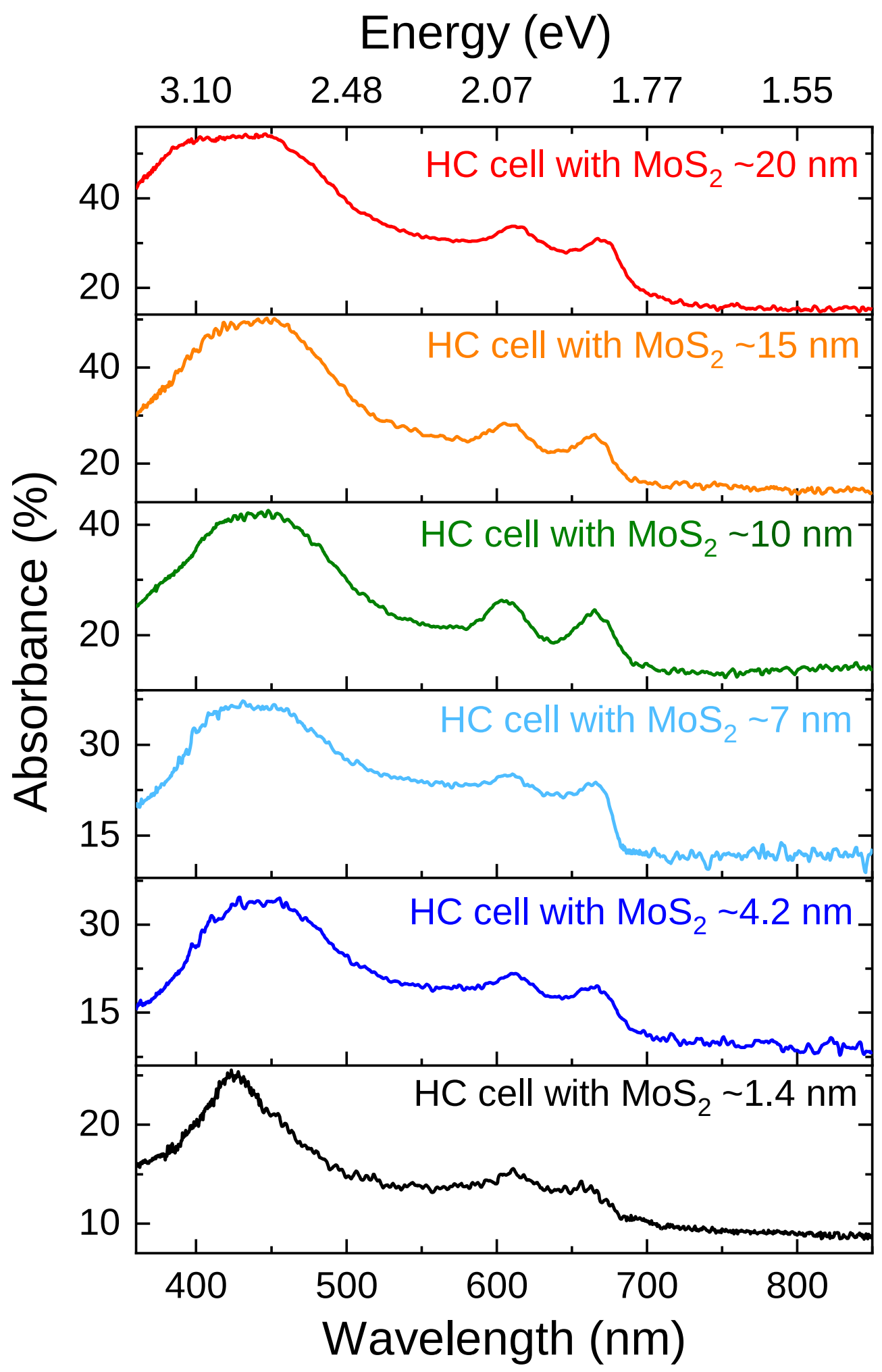

**Extended Data Fig. 1| Optical absorbance of $MoS_2$/Gr/$TiO_X$/Ti cells.** Optical absorbance of cells with various $MoS_2$ thicknesses. Reflectance and transmittance of the cells on quartz substrate were measured to calculate the absorbance.

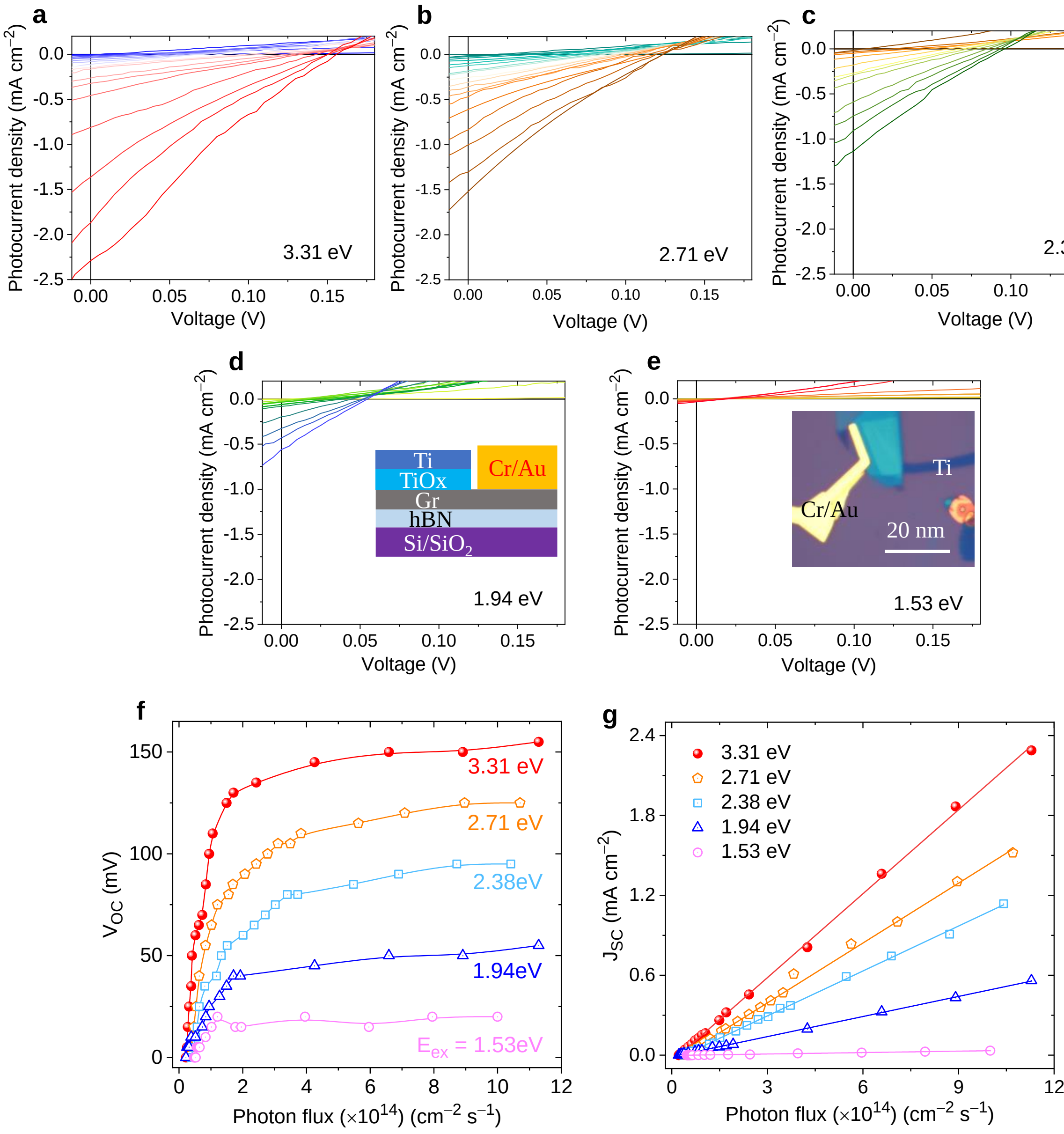

**Extended Data Fig. 2| Hot carrier-driven photovoltaic Gr/TiOx cell. a-e**, Photocurrent density vs voltage with various photon densities (up to ~$1.2\times10^{15}$ $cm^{-2}$ $s^{-1}$) at different photon energies (3.31 eV, 2.71 eV, 2.38 eV, 1.94 eV and 1.53 eV). Schematic representation of Gr/$TiO_X$/Ti cell on hBN substrate is shown in inset of **d**, while the corresponding optical micrograph is depicted in inset of **e. f**, Open circuit voltage ($V_{OC}$) and **g**, short circuit current density ($J_{SC}$) as a function of photon density for five different energies, clearly revealing the HC extraction from Gr via resonant tunneling through BRT layer.

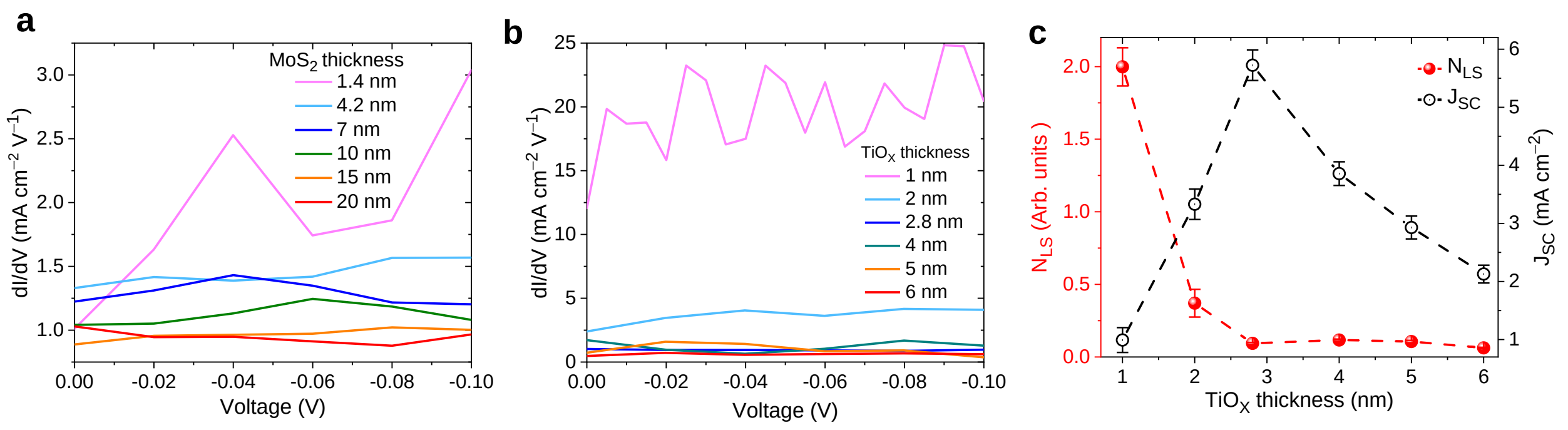

**Extended Data Fig. 3| Leakage states varying with $MoS_2$ and $TiO_X$ thickness.** 1$^{st}$ derivative of photocurrent with respect to the voltage (*dI/dV*) with various **a**, $MoS_2$ film and **b**, BRT layer thicknesses at negative voltage from 0 to –0.1 V. **c**, Number of leakage states ($N_{LS}$) in the BRT layer varying with $TiO_X$ thicknesses. Leakage current dominates at low BRT layer thickness (1 nm) and rapidly falls with increasing $TiO_X$ layer thickness over 2.8 nm.

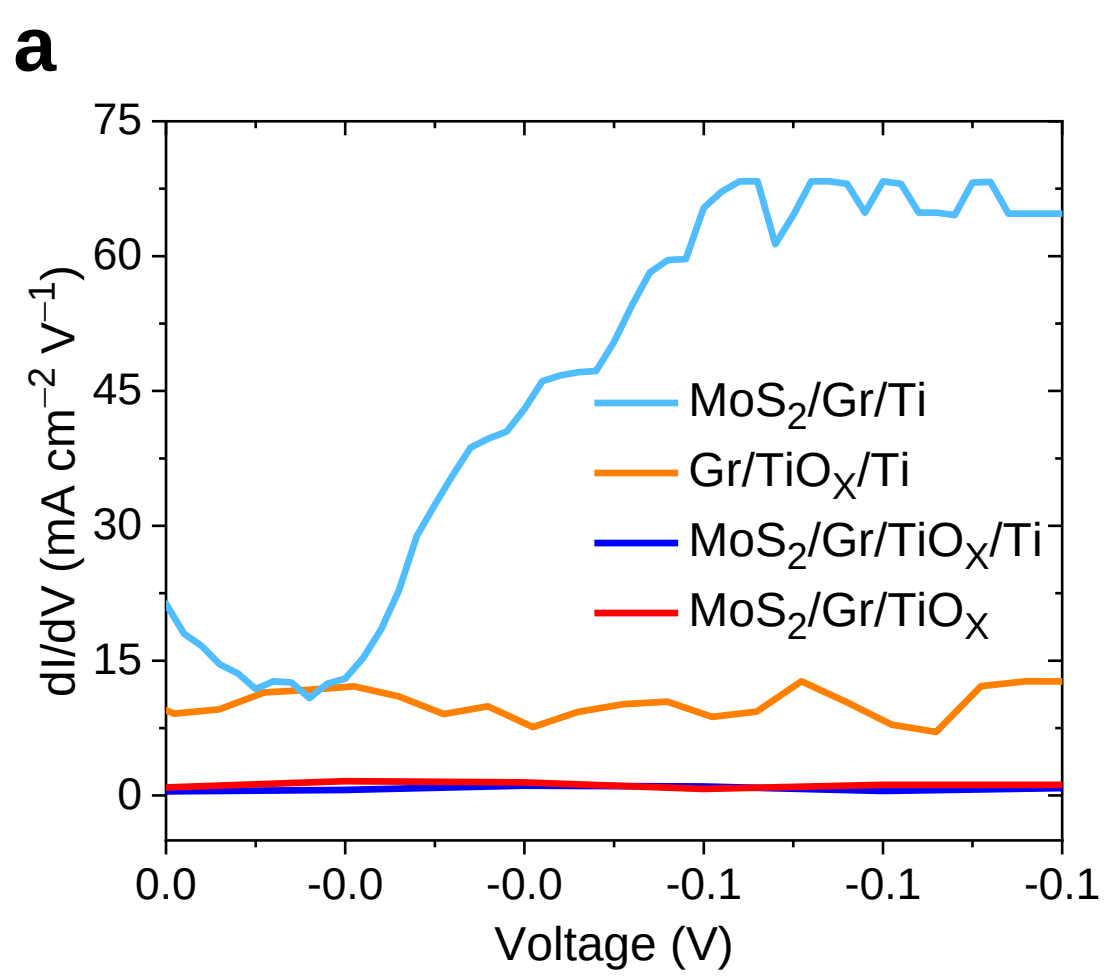

**Extended Data Fig. 4| Leakage and tunneling states in various cells.** *dI/dV* for **a**, leakage states at negative voltage from 0 to –0.1 V and **b**, tunneling states at positive voltage from 0 to 1 V.